\documentclass[aps,prl,reprint,amsmath,amssymb,showpacs,superscriptaddress]{revtex4-1}
\usepackage{graphicx}
\usepackage{dcolumn}
\usepackage{bm}

\usepackage{hyperref}

\begin{document}


\title{A microscopic resolution of the chiral conundrum with crossing twin bands in $^{106}$Ag }

\author{P. W. Zhao 
}
\email{pwzhao@pku.edu.cn}
\affiliation{State Key Laboratory of Nuclear Physics and Technology, School of Physics, Peking University \\ 
Beijing 100871, China}

\author{Y. K. Wang 
}
\affiliation{State Key Laboratory of Nuclear Physics and Technology, School of Physics, Peking University \\ 
Beijing 100871, China}

\author{Q. B. Chen
}
\affiliation{Physik-Department, Technische Universit\"{a}t M\"{u}nchen, D-85747 Garching, Germany}


\begin{abstract}
The nuclear chiral conundrum with crossing twin bands is investigated with three-dimensional tilted axis cranking covariant density functional theory in a fully self-consistent and microscopic way. 
The energy spectra and electromagnetic transition strengths for bands 1 and 2 in $^{106}$Ag are well reproduced with two distinct configurations with two and four quasiparticles, respectively.
For the four-quasiparticle configuration, a chiral vibrational band on top of band 2 is expected due to the soft Routhian curves. 
Therefore, it provides a microscopic and solid solution for the chiral conundrum in $^{106}$Ag. It also paves the way for understanding similar chiral structure in other nuclei in the future.. 
\end{abstract}



\maketitle



\section{Introduction}
Chirality is a well-known phenomenon in many fields, such as chemistry, biology, molecular and particle physics. 
In nuclear physics, chirality was originally suggested by Frauendorf and Meng in 1997~\cite{Frauendorf1997Nucl.Phys.A131}. 
It represents a novel feature of rotating triaxial nuclei, where three angular momentum vectors in the intrinsic frame may couple to each other in either a left- or right-handed mode. 
The two modes differ from each other by their intrinsic chirality, and are thus connected by the chiral operator $TR(\pi)$ that combines time reversal $T$ and spatial rotation by $\pi$. 

In the laboratory frame, the restoration of the broken intrinsic chiral symmetry 
gives rise to the so-called chiral doublet bands, which consist of a pair of nearly-degenerate $\Delta I=1$ bands (for reviews see Refs.~\cite{Frauendorf2001Rev.Mod.Phys.463,Meng2010JPhysG.37.64025}).  
The existence of chiral doublet bands have been reported experimentally in the $A\sim 80$, 100, 130, and 190 mass regions of the nuclear chart; see e.g., Refs.~\cite{Starosta2001Phys.Rev.Lett.971,Zhu2003Phys.Rev.Lett.132501,Vaman2004Phys.Rev.Lett.32501,Grodner2006Phys.Rev.Lett.172501,Joshi2007Phys.Rev.Lett.102501,Mukhopadhyay2007Phys.Rev.Lett.172501,Ayangeakaa2013Phys.Rev.Lett.172504,Kuti2014Phys.Rev.Lett.32501,Tonev2014Phys.Rev.Lett.52501,Liu2016Phys.Rev.Lett.112501,Grodner2018Phys.Rev.Lett.22502,Xiong2019At.DataNucl.DataTables193}. 
In most cases, the doublet bands are separated in energy at low spins, while they approach each other with increasing spin and become approximately degenerate above a critical spin. 
This feature has been understood as a transition from a chiral vibrational mode at low spins to the static chiral mode above a critical spin~\cite{Olbratowski2004Phys.Rev.Lett.52501,Mukhopadhyay2007Phys.Rev.Lett.172501,Qi2009Phys.Lett.B175}, which is a consequence of quantum tunneling between the intrinsic left- and right-handed chiral solutions.  

Although the energies of chiral doublet bands are close to each other, a crossing between the twin bands is rarely observed. 
Two most famous examples are in the nuclei $^{134}$Pr~\cite{Starosta2001Phys.Rev.Lett.971} and $^{106}$Ag~\cite{Joshi2007Phys.Rev.Lett.102501}. 
Such crossing bands have triggered extensive investigations since they may provide an ideal test of the onset of static chirality~\cite{Starosta2001Phys.Rev.Lett.971,Joshi2007Phys.Rev.Lett.102501,Koike2004Phys.Rev.Lett.172502,Tonev2006Phys.Rev.Lett.52501,Petrache2006Phys.Rev.Lett.112502,Lieder2014Phys.Rev.Lett.202502,Rather2014Phys.Rev.Lett.202503}. 
Indeed, for some time, $^{134}$Pr was regarded as the best example of nuclear chirality. 
However, this conclusion is not supported by the subsequent experimental measurements of the electromagnetic transition rates~\cite{Tonev2006Phys.Rev.Lett.52501}. 
In particular, the measured in-band $B(E2)$ values for the candidate chiral partner bands show large differences, and this is not in harmony with the picture of a good static chirality, which requires similar electromagnetic transition rates for the twin bands~\cite{Meng2010JPhysG.37.64025}. 

The nature of the other famous case of crossing bands in $^{106}$Ag is even more elusive. 
In Ref.~\cite{Joshi2007Phys.Rev.Lett.102501}, the excited partner band was explained in terms of an axial shape resulting from a novel shape transformation induced by chiral vibration from the triaxial yrast band to the axial excited partner band. 
Nevertheless, two recent independent lifetime measurements~\cite{Lieder2014Phys.Rev.Lett.202502,Rather2014Phys.Rev.Lett.202503} reported similar $B(E2)$ and $B(M1)$ values for the partner crossing bands; reflecting that the two bands may built on similar nuclear shapes.  
Therefore, the characterization of the crossing bands and the corresponding chiral manifestation is still an open question. 

Another important feature of the crossing bands is the observation of the third band lying only slightly higher than the two existing crossing bands in $^{134}$Pr and $^{106}$Ag ~\cite{Timar2011Phys.Rev.C44302,Lieder2014Phys.Rev.Lett.202502}.
In particular, for $^{106}$Ag, the electromagnetic transition rates for the third band have also been measured~\cite{Lieder2014Phys.Rev.Lett.202502}. 
There are certain signs from the systematic behaviors of the data indicating that the third band together with one of the crossing bands might form a pair of chiral doublet bands. 
This interpretation is also consistent with the calculated results given by the particle rotor model (PRM), which however, is a phenomenological model and is adjusted to the data in one way or another~\cite{Lieder2014Phys.Rev.Lett.202502}. 

Therefore, to explore the mysteries associated with the chiral manifestation of the crossing bands, it is highly desirable to perform a self-consistent and microscopic investigation. 
Such calculations are more challenging, but they are nowadays feasible in the framework of density functional theories (DFTs). 
The DFTs provide a fully self-consistent mean field for nucleons, which depends entirely on a universal energy density functional for the entire nuclide chart and, thus, would provide a thorough understanding of the chiral conundrum.  

Covariant DFT exploits basic properties of QCD at low energies, in particular, the presence of symmetries and the separation of scales~\cite{Lalazissis2004}. 
It provides a consistent treatment of the spin degrees of freedom, includes the complex interplay between the large Lorentz scalar and vector self-energies induced at the QCD level~\cite{Cohen1992Phys.Rev.C1881}, and naturally provides the nuclear currents induced by the spatial parts of the vector self-energies, which play an essential role in rotating nuclei. 
To describe nuclear rotation, covariant DFT has been extended with the cranking method~\cite{Koepf1989Nucl.Phys.A61,Madokoro2000Phys.Rev.C61301,Peng2008Phys.Rev.C24313,Zhao2011Phys.Lett.B181,Zhao2017Phys.Lett.B1}, and this has provided a satisfactory description of rotational excited states all over the periodic table and has demonstrated high predictive power~\cite{Afanasjev1999Phys.Rep.1,Meng2013Front.Phys.55,Meng2016PhysicaScripta53008,Zhao2011Phys.Rev.Lett.122501,Zhao2015Phys.Rev.Lett.22501,Zhao2018Int.J.Mod.Phys.E1830007}. 
For nuclear chirality, in particular, the recently developed three-dimensional tilted axis cranking (3DTAC) approach based on covariant DFT has been successfully applied for describing the multiple chirality in $^{106}$Rh~\cite{Zhao2017Phys.Lett.B1}. 

In the present paper, the chiral conundrum associated with the crossing partner bands will be investigated with the 3DTAC approach based on covariant DFT by taking the nucleus $^{106}$Ag as an example. 
The calculations are fully self-consistent and microscopic, and are free of any readjustment of parameters to the observed band structure in $^{106}$Ag. 
Therefore, they provide a test for the energy density functional applying to the chiral conundrum in  $^{106}$Ag.

\section{Theoretical Framework}
Covariant DFT starts from a Lagrangian, and the corresponding Kohn-Sham equations have the form of a Dirac equation with effective fields $S(\bm{r})$ and $V^\mu(\bm{r})$ derived from this Lagrangian~\cite{Ring1996Prog.Part.Nucl.Phys.193,Vretenar2005Phys.Rep.101,Meng2006Prog.Part.Nucl.Phys.470,Niksic2011Prog.Part.Nucl.Phys.519,Meng2015}. In the 3DTAC method~\cite{Zhao2017Phys.Lett.B1}, these fields are triaxially deformed, and the calculations are carried out in the intrinsic frame rotating with a constant angular velocity vector $\bm{\omega}$, pointing in an arbitrary direction in space:
\begin{equation}\label{Eq.Dirac}
	\left[\bm{\alpha}\cdot(\bm{p}-\bm{V}) + \beta(m+S) + V - \bm{\omega}\cdot\hat{\bm{J}} \right]\psi_k = \epsilon_k\psi_k.
\end{equation}
Here, $\hat{\bm{J}}$ is the total angular momentum of the nucleon spinors, and the fields $S$ and $V^\mu$ are connected in a self-consistent way to the nucleon densities and current distributions, which are obtained from the single-nucleon spinors $\psi_k$~\cite{Zhao2012Phys.Rev.C54310,Meng2013Front.Phys.55,Zhao2018Int.J.Mod.Phys.E1830007}. 
The iterative solution of these equations yields single-particle energies, expectation values for the three components $\langle \hat{J}_i\rangle$ of the angular momentum, total energies, quadrupole moments, transition probabilities, etc. 
The magnitude of the angular velocity $\bm{\omega}$ is connected to the angular momentum quantum number $I$ by the semiclassical relation $\langle \hat{\bm{J}}\rangle\cdot\langle\hat{\bm{J}} \rangle=I(I+1)$, and its orientation is determined by minimizing the total Routhian self-consistently. 

Pairing correlations are considered by solving the tilted axis cranking relativistic Hartree Bogoliubov (TAC-RHB) equations in the framework of superfluid covariant DFT~\cite{Zhao2015Phys.Rev.C34319,Wang2017Phys.Rev.C54324}. 
The TAC-RHB model achieves a unified and self-consistent treatment of the mean fields, which include long range particle-hole (\emph{ph}) correlations, and the pairing field which sums up the particle-particle (\emph{pp}) correlations. For details on the TAC-RHB method, one can see Refs.~\cite{Zhao2015Phys.Rev.C34319,Wang2017Phys.Rev.C54324,Wang2018Phys.Rev.C64321}. 

In this work, the point-coupling Lagrangian PC-PK1~\cite{Zhao2010Phys.Rev.C54319} is adopted in the \emph{ph} channel, and a finite-range separable pairing force~\cite{Tian2009Phys.Lett.B44} is used in the \emph{pp} channel. 
The scaling factor of the pairing strength is taken from Ref.~\cite{Agbemava2014Phys.Rev.C54320} according to a global analysis of nuclear ground-state properties. The calculations are free of additional parameters. 
The Dirac equation [Eq.~(\ref{Eq.Dirac})] is solved in a three-dimensional Cartesian harmonic oscillator basis with 10 major shells.
It has been checked that the total energy at rotational frequency $\hbar\omega=$ 0.25 MeV changes only by 0.04\% with 12 major shells, and the corresponding obtained deformation is barely changed. 

The present study focuses on the odd-odd nucleus $^{106}$Ag. In the latest experiment of this nucleus~\cite{Lieder2014Phys.Rev.Lett.202502}, three close-lying bands of negative parity were reported. Most previous works~\cite{Joshi2007Phys.Rev.Lett.102501,Rather2014Phys.Rev.Lett.202503,Lieder2014Phys.Rev.Lett.202502} have assumed that band 1 corresponds to the two-quasiparticle configuration $\pi g_{9/2}\otimes\nu h_{11/2}$, where a quasi-proton in $g_{9/2}$ shell is coupled with a quasi-neutron in $h_{11/2}$ shell. In Ref.~\cite{Lieder2014Phys.Rev.Lett.202502}, a four-quasiparticle 
configuration, $\pi g_{9/2}\otimes\nu h_{11/2}(gd)^2$, where a pair of quasi-neutrons in the low-$j$ $(g_{7/2} d_{5/2})$ shell are aligned, is assumed to be the configuration of bands 2 and 3. In this work, we have carried out the self-consistent 3DTAC calculations with both configurations in the framework of covariant DFT.
The configurations have been identified by expanding the single-particle orbitals on a set of spherical harmonic oscillator basis. They are fixed during the iterative solution of the Dirac equation by calculating the maximum overlap between each block orbital at two successive iterations~\cite{Peng2008Phys.Rev.C24313,Zhao2015Phys.Rev.C34319}.

\section{Results and discussion}
\begin{figure}[!htbp]
\centering
\includegraphics[width=8.0cm]{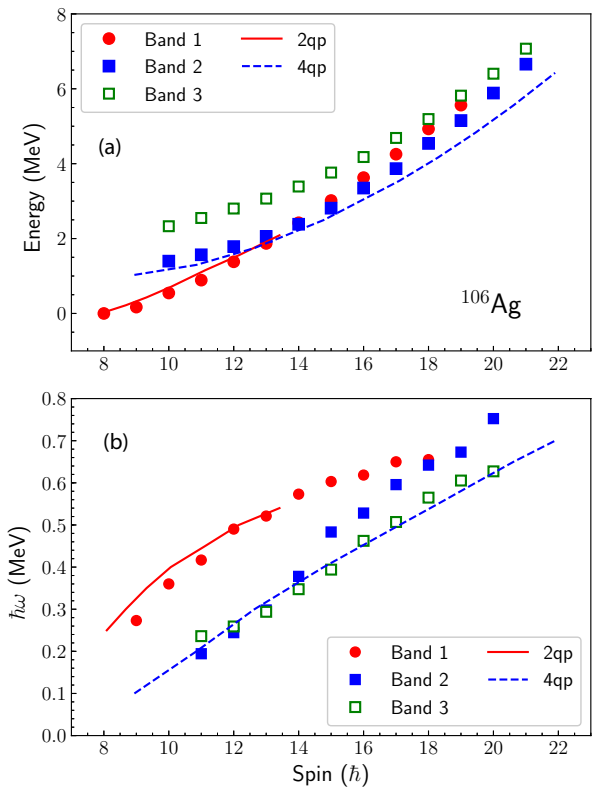}
\caption{(color online.) Calculated rotational excitation energies (top) and rotational frequencies (bottom) with the two- and four- quasiparticle configurations as a function of the angular momentum in comparison with data~\cite{Lieder2014Phys.Rev.Lett.202502}. The excitation energies are renormalized to the bandhead.}
\label{fig1}
\end{figure}

In Fig.~\ref{fig1}, the calculated excitation energies and the rotational frequencies $\hbar\omega$ are compared with data~\cite{Lieder2014Phys.Rev.Lett.202502}. 
The calculated results with the two- and four- quasiparticle configurations can reproduce very well the data of bands 1 and 2, respectively.  
This demonstrates clearly that the crossing between bands 1 and 2 are caused by the different configurations. 
This is consistent with the assumptions adopted in the previous phenomenological PRM calculations~\cite{Lieder2014Phys.Rev.Lett.202502}. 
In the present work, however, the configuration assignment is confirmed solidly in a microscopic and self-consistent framework of covariant DFT. 
In particular, the energy separation between the bandheads of bands 1 and 2 is reproduced very well, but this cannot be achieved in a phenomenological PRM. 
Moreover, the calculated energy difference between the bandhead of band 1 and the ground state is 0.75 MeV, which is very close to the experimental value 0.87 MeV.

We found that pairing correlations play a significant role in the two-quasiparticle band, while they become negligible in the four-quasiparticle band due to the alignment of two more quasiparticles.   
For the two-quasiparticle configuration, convergent results can be obtained only up to around $14\hbar$, where the alignment of band 1 indicates an onset of band crossing~\cite{Lieder2014Phys.Rev.Lett.202502}. 
It is well-known that cranking approaches are not appropriate 
for describing band crossings~\cite{Hamamoto1976Nucl.Phys.A15}. 

There is no proper configuration obtained for band 3 in the present calculations.
Considering the fact that bands 2 and 3 are lying close to each other with similar quasiparticle alignments~\cite{Lieder2014Phys.Rev.Lett.202502}, it  indicates that band 3 might be a chiral partner band of band 2. 
At the present mean-field level, it does not take into account either the chiral vibrations nor the tunneling between the left- and right-handed sectors. 
Therefore, the energy splitting between bands 2 and 3 cannot be calculated. 
Further extensions going beyond the mean field by using, for instance, the methods of the random phase approximation~\cite{Mukhopadhyay2007Phys.Rev.Lett.172501} or the collective Hamiltonian method~\cite{Chen2013Phys.Rev.C24314,Chen2016Phys.Rev.C44301} will be required for this purpose in the framework of DFTs. 

\begin{figure}[!htbp]
\centering
\includegraphics[width=7.0cm]{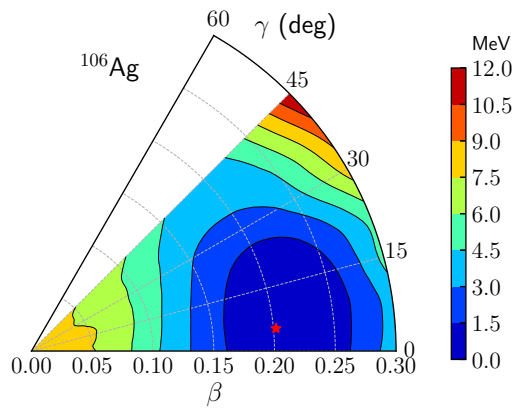}
\caption{(color online.) Potential energy surface of $^{106}$Ag in the $\beta$-$\gamma$ deformation plane for the configuration $\pi g_{9/2}\otimes\nu h_{11/2}(gd)^2$ at the rotational frequency $\hbar\omega = 0.25$ MeV. The star denotes the position of the minimum energy. }
\label{fig2} 
\end{figure}

However, to justify the chiral nature of bands 2 and 3, one can first check the magnitude of triaxial deformation for the four-quasiparticle configuration, which is obtained self-consistently in this work. 
In Fig.~\ref{fig2}, the potential energy surface of $^{106}$Ag at the rotational frequency $\hbar\omega = 0.25$ MeV is shown with the configuration fixed to be the four-quasiparticle one. 
Although the triaxial deformation is only $\gamma\sim5^\circ$ at the energy minimum, the potential energy surface is rather soft in the triaxial direction; the energy rise is less than 1.5 MeV with the triaxial deformation reaching $\gamma\sim25^\circ$. 
Considering the fact that the four-quasiparticle configuration contains high-$j$  protons and neutrons in the $g_{9/2}$ and $h_{11/2}$ shells, respectively,  a partner band of chiral vibrational mode is very likely to be built on top of band 2.   

To examine the possible presence of chiral vibration, it is crucial to check the calculated orientation angles $\theta$ and $\phi$ of the total angular momentum $\bm{J}$ in the intrinsic frame. 
Here, $\theta$ is the angle between the angular momentum and the long axis, and $\phi$ the angle between the angular momentum projection onto the intermediate-short plane and the short axis~\cite{Zhao2017Phys.Lett.B1}.
In the present calculations with the four-quasiparticle configuration, the polar angle $\theta$ varies from $46^\circ$ to $69^\circ$ driven by the increasing rotational frequency, while the azimuth angle $\phi$ vanishes at all rotational frequencies. 
This provides a planar rotation, where the angular momentum lies in the plane of short and long axes. 

It should be noted that 3DTAC gives only the classical orientation, around which the angular momentum $\bm{J}$ can execute a quantal motion. 
In the planar rotation ($\phi = 0$), the angular momentum vector $\bm{J}$ could oscillate around the planar equilibrium into the left- ($\phi < 0$) and right-handed ($\phi > 0$) sectors, and this leads to the so-called chiral vibration~\cite{Starosta2001Phys.Rev.Lett.971}. 
As a result, two separate bands are expected to be observed, corresponding to the first two vibrational states. 
Therefore, the experimental observation of chiral vibrations requires a relatively low vibrational energy, which in turn requires that the Routhian rises slowly along the $\phi$ degree of freedom~\cite{Chen2013Phys.Rev.C24314}.  

\begin{figure}[!htbp]
\centering
\includegraphics[width=8.5cm]{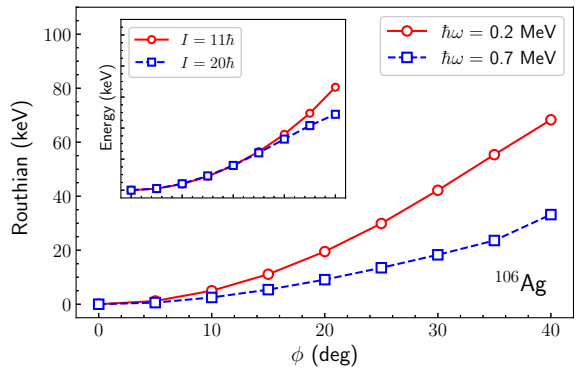}
\caption{(color online.) Total Routhian curves for the configuration $\pi g_{9/2}\otimes\nu h_{11/2}(gd)^2$ at rotational frequencies $\hbar\omega = $ 0.2 and 0.7 MeV as functions of the azimuth angle $\phi_\omega$ of the angular velocity $\bm{\omega}$, which are determined by minimizing the total Routhian with respect to the polar angle $\theta_\omega$ for each given value of $\phi_\omega$. The two Routhian curves are respectively renormalized to their minima at each rotational frequency. Inset: similar to the Routhian curves but for the total energy curves at spin $I = 11\hbar$ and $I=20\hbar$ as functions of the azimuth angle $\phi_I$ of the spin. } 
\label{fig3} 
\end{figure}

This can be seen from Fig.~\ref{fig3}, where the total Routhian curves are shown as functions of $\phi_\omega$ for the four-quasiparticle configuration at rotational frequencies $\hbar\omega = $ 0.2 and 0.7 MeV.
Here, the azimuth angle $\phi_\omega$ and the polar angle $\theta_\omega$ are used to represent the orientation of the angular velocity $\bm{\omega}$. 
The total Routhian curves are determined by minimizing the total Routhian with respect to $\theta_\omega$ for each given value of $\phi_\omega$. 
The validity of the Kerman-Onishi conditions~\cite{Kerman1981Nucl.Phys.179} has been checked similar to Ref.~\cite{Shi2013Phys.Rev.C34311}, and it is found that the Kerman-Onishi conditions are satisfied with a high precision, which is sufficient for determining the observables in the present 3DTAC-CDFT calculations.

It is seen in Fig.~\ref{fig3} that for both rotational frequencies $\hbar\omega = $ 0.2 and 0.7 MeV, the Routhian grows very slowly with the increasing $\phi_\omega$; rising only several tens of keV from $\phi_\omega = 0^\circ$ to $40^\circ$. 
Similar behavior can be also seen for the total energies which grows slowly with the increasing $\phi_I$ at spin $I = 11\hbar$ and $I=20\hbar$.  
This indicates that the chiral vibration around the planar equilibrium into the left- and right-handed sectors should be substantial, and a pair of chiral vibration bands can be generated. 
Moreover, the present calculations demonstrate that the Routhian curve becomes softer with respect to the $\phi$ direction at higher rotational frequencies; indicating that the chiral vibrational energies are smaller at high angular momentum. 
Therefore, one can expect that the observed energy separation between the chiral twin bands would be reduced with increasing spin, and this is indeed consistent with the experimental data of bands 2 and 3 as shown in Fig.~\ref{fig1}. 

\begin{figure}[!htbp]
\centering 
\includegraphics[width=8.0cm]{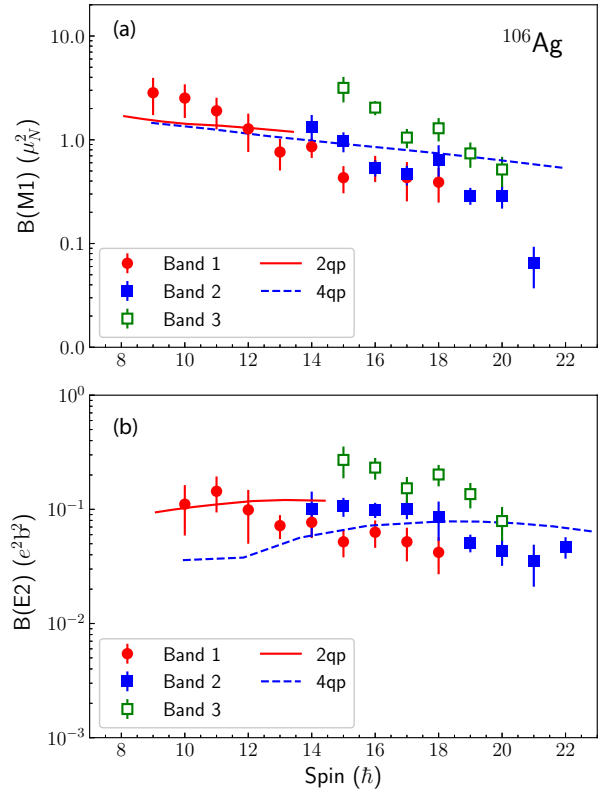}
\caption{(color online.) Calculated $M1$ (top) and $E2$ (bottom) transition probabilities with the two- and four- quasiparticle configurations as a function of the angular momentum in comparison with the data~\cite{Lieder2014Phys.Rev.Lett.202502}. }
\label{fig4} 
\end{figure}

The $B(M1)$ and $B(E2)$ transition probabilities can be calculated in the semiclassical approximation from the magnetic and quadrupole moments, respectively~\cite{Frauendorf1997Nucl.Phys.A131}. 
Here, the magnetic moments are derived from the relativistic electromagnetic current operator as in Ref.~\cite{Zhao2017Phys.Lett.B1}. 
In Fig.~\ref{fig4} the calculated $B(M1)$ and $B(E2)$ values with the two- and four- quasiparticle configurations are shown as a function of the angular momentum in comparison with the data from the latest lifetime measurements~\cite{Lieder2014Phys.Rev.Lett.202502}.

The experimental electromagnetic transition rates for bands 1 and 2 are well reproduced by the calculated results with the two- and four- quasiparticle  configurations, respectively. 
This provides a further strong support for the present configuration assignment of bands 1 and 2.   
For both bands, it is found that the deformation changes only slightly along the band, so the corresponding $B(E2)$ values are roughly constant. 
However, the $B(M1)$ values decrease smoothly along the band because of the so-called shears mechanism~\cite{Frauendorf2001Rev.Mod.Phys.463}, i.e., the gradual close of the neutron and proton angular momentum vectors. 

Note that the configurations of bands 1 and 2 differ only slightly by two quasi-neutrons in the low-$j$ shells of $g_{7/2}$ and $d_{5/2}$, which  influence gently the deformation parameters $(\beta,\gamma)$ and the rotational orientation $(\theta,\phi)$. 
As a result, the calculated electromagnetic transition properties of bands 1 and 2 are very close to each other. 
This explains nicely the behaviors of the observed data, and demonstrates clearly that bands 1 and 2 do not form a pair of chiral partner bands.  

\section{Summary}
In summary,  a fully self-consistent and microscopic investigation for the chiral conundrum associated with the crossing partner bands in $^{106}$Ag has been carried out with the 3DTAC approach based on covariant DFT.
The calculated energy spectra and electromagnetic transition probabilities with two distinct configurations $\pi g_{9/2}\otimes\nu h_{11/2}$ and $\pi g_{9/2}\otimes\nu h_{11/2}(gd)^2$ are in good agreement with the corresponding data of bands 1 and 2. 
For the latter configuration, it is found that the  potential energy surface is rather soft with respect to the triaxial degree of freedom.
Moreover, due to the soft Routhian curves, the chiral vibration around the planar equilibrium into the left- and right-handed sectors can be substantial. 
A pair of chiral vibration bands are thus expected, and this is consistent with the latest observations~\cite{Lieder2014Phys.Rev.Lett.202502}. 
Therefore, the present work provides a microscopic and solid solution for the chiral conundrum in $^{106}$Ag. 
It also paves the way for understanding similar chiral structure in other nuclei in the future.

\begin{acknowledgments}
The authors thank J. Meng, and S. Q. Zhang for helpful discussions. 
This work was supported in part by the National Key R\&D Program of China (Contract No. 2018YFA0404400 and No. 2017YFE0116700), the National Natural Science Foundation of China (Grants No. 11775026 and No. 11875075), the Deutsche Forschungsgemeinschaft (DFG) and National Natural Science Foundation of China (NSFC) through funds provided to the Sino-German CRC 110 “Symmetries and the Emergence of Structure in QCD” (DFG Grant No. TRR110 and NSFC Grant No. 11621131001), and the Laboratory Computing Resource Center at Argonne National Laboratory.
\end{acknowledgments}


%

\end{document}